\documentstyle[preprint,aps,eqsecnum]{revtex}
\draft
\begin{document}
\title{On the Mechanism of Pinning in Phase-Separating Polymer Blends}
\author{Claudio Castellano\footnote{Permanent address: Dipartimento di Scienze
Fisiche, Universit\`a degli studi di Napoli, Pad. 19, Mostra d'Oltremare,
Napoli
80125, Italy. Internet address: castellano@na.infn.it} and Sharon C.
Glotzer\footnote{Internet address: glotzer@ctcms.nist.gov}}
\address {Center for Theoretical and Computational Materials Science, and
Polymers Division, National Institute of Standards and Technology,
Gaithersburg, MD 20899}

\date{\today}

\maketitle

\begin{abstract}
We re-explore the kinetics of spinodal decomposition in off-critical
polymer blends through numerical simulations of the Cahn-Hilliard
equation with the Flory-Huggins-De Gennes free energy functional. Even
in the absence of thermal noise, the solution of the discretized
equation of motion shows coarsening in the late stages of spinodal
decomposition without evidence of pinning, regardless of the relative
concentration of the blend components.  This suggests this free energy
functional is not sufficient to describe the physics responsible for
pinning in real blends.
\end{abstract}

\vfill\eject

%\begin{narrowtext}
\section{Introduction}

Experiments on spinodal decomposition in polymer blends show that the
coarsening process may sometimes slow dramatically or even cease
before reaching equilibrium \cite{hasegawa88} - \cite{lauger94}.  In
these systems, spinodal decomposition --- which is the process by
which a thermodynamically unstable mixture demixes to a stable,
phase-separated equilibrium state \cite{cahn} - \cite{glotzer95} ---
proceeds normally for some time following a quench to the unstable
region ($T<T_s$), and then stops. A break-up of the characteristic,
interconnected pattern is observed to precede this pinning phenomenon.
The nonequilibrium, microphase-separated blend has been observed to
remain in this pinned state over an appreciable time scale where
little domain growth occurs.  The eventual breakup of the evolving
morphology into separated droplets is a natural consequence of the
asymmetric composition in an off-critical blend \cite{hayward87};
nevertheless, it may also occur in near-critical blends due to other
forces (e.g. gravity). Polymer blends in which pinning has recently
been observed for off-critical composition include X-7G/poly(ethylene
teraphthalate) (a liquid crystalline polymer/homopolymer blend)
\cite{hasegawa88}, poly(styrene-ran-butadiene)/polybutadiene
\cite{hashimoto92}, poly(styrene-ran-butadiene)/polyisoprene
\cite{hashimoto92}, and polybutadiene/polyisoprene \cite{lauger94}.

The specific mechanism responsible for pinning in these blends is
poorly understood, and is currently a topic of considerable
discussion.  While there is general agreement that growth stops soon
after the breakup into separated ``droplets'' or ``clusters'' (a so-called
``percolation-to-cluster transition'' \cite{hashimoto92,hayward87}),
the mechanism that prevents further coarsening of disconnected domains
remains to be clarified.  One intriguing scenario points to an
entropic barrier as the reason for the observed arrested growth of
off-critical phase separating blends.  Kotnis and Muthukumar (KM)
\cite{kotnis92} have suggested that due to the connectivity of the
chains and the reduced conformational entropy near domain interfaces
\cite{helfand}, the usual evaporation-condensation mechanism
of coarsening \cite{lifshitz61}
observed in small-molecule mixtures is suppressed in polymer blends,
and instead coarsening occurs via parallel transport of chains along
the interface \cite{huse86}.  Consequently, KM postulate that if the
concentration of the minority-rich phase becomes smaller than the
percolation threshold, the parallel coarsening mechanism will be
inhibited and the clusters will ``freeze'' after an initial growth
period.

Hashimoto and coworkers have instead postulated that the enthalpy of
mixing, rather than the entropy, provides the barrier to further
coarsening following the percolation-to-cluster transition
\cite{hashimoto92}.  They argue that the increase in enthalpy of
mixing suffered upon removing a chain of species $A$ and degree of
polymerization $N$ from the surface of an $A$-rich domain is $\Delta
H_{\rm mix} \propto \chi N k_BT$, where $\chi$ is the Flory
interaction parameter, $k_B$ is Boltzmann's constant, and $T$ is
temperature.  In the strong segregation limit $\chi N \gg 1$, and thus
evaporation of the chain from the domain surface, which would occur
with a Boltzmann probability proportional to $\exp (-\Delta H_{\rm
mix}/ k_BT)$, is highly unfavorable.  Thus, when the parallel
transport mechanism is eliminated by the breakup into droplets,
coarsening ceases.

In this paper, we re-explore the kinetics of spinodal decomposition in
off-critical polymer blends described by the Flory-Huggins-De Gennes
(FHDG) free energy functional, through numerical simulations of the
Cahn-Hilliard (CH) equation. In Sec.~\ref{sec:theory}, we discuss the
CH-FHDG equation and the origin of the concentration-dependent square
gradient coefficient that has been proposed by KM to cause pinning in
off-critical blends.  The discretization and numerical integration
scheme used to solve this equation, and our numerical results, are
presented in Sec.~\ref{sec:solution} and discussed in
Sec.~\ref{sec:discussion}. Finally, a summary of our main conclusions,
and speculations on possible mechanisms of pinning in blends, is
discussed in Sec.~\ref{sec:conclusion}.

\section{Theoretical Model}
\label{sec:theory}

Model blends are typically described by the Flory-Huggins-De Gennes
free energy functional \cite{degennes80,pincus81,binder83}:
\begin{eqnarray}
{F\{\phi({\bf r})\} \over k_B T}= \int d^3r \left[{f_{FH}(\phi ({\bf
r}))\over k_BT} + \kappa(\phi)(\nabla \phi)^2 \right],
\label{eq:fhdg}
\end{eqnarray}
with
\begin{eqnarray}
\kappa(\phi) = {\sigma_a ^2 \over 36 \phi} + {\sigma_b^2
\over 36 (1-\phi)} + \chi \lambda ^2,
\label{eq:kappa}
\end{eqnarray}
where $\phi({\bf r})$ is the local concentration of component $A$ (so
that 1-$\phi$ is the concentration of component $B$), $\sigma_A$ and
$\sigma_B$ are the Kuhn lengths of the two species, $\lambda$ is an
effective interaction distance between monomers, and the Flory-Huggins
free energy is \cite{flory,huggins}
\begin{eqnarray}
{{f_{FH}(\phi)}\over k_BT} = {\phi \over N_A} \ln \phi + {(1-\phi) \over N_B}
\ln (1-\phi) + \chi \phi (1-\phi),
\label{eq:fh}
\end{eqnarray}
where $N_A$ ($N_B$) is the degree of polymerization of chains $A$
($B$).  Whereas in small molecule mixtures the square gradient
coefficient is enthalpic (arising from short range interactions
between molecules) and independent of the local concentration
\cite{fredrickson92}, De Gennes proposed that the connectivity of
polymer molecules in inhomogeneous blends manifests itself through an
additional, concentration-dependent contribution to the square
gradient coefficient $\kappa(\phi)$ \cite{degennes80}.  The expression
for $\kappa(\phi)$ in Eq.~\ref{eq:kappa} was derived to be consistent
with the random phase approximation result for the inverse structure
factor of an incompressible polymer blend
\cite{binder83,edwards66,akcasu},
\begin{eqnarray}
S^{-1}({\bf q}) = {1\over {N_A\phi_o D(q^2R_A^2)}} +
{1\over {N_B(1-\phi_o) D(q^2R_B^2)}} - 2\chi.
\end{eqnarray}
Here $R_i^2$ is the average square radius of gyration of species $i$,
$\phi_o$ is the average value of the concentration
and the Debye function is $D(x) = 2[x-1+ e^{-x}]/x^2$, with $x \equiv
q^2R_i^2$.  In the weak segregation limit, the interfacial width
is much larger than the chain dimensions \cite{fredrickson92}, so that
the length scales of interest are larger than $R_i$ ($q^2R_i^2 \ll
1$), and thus the Debye function may be approximated by $D^{-1}(x) =
1+x/3+O(x^2)$.  The square gradient coefficient in Eq.~\ref{eq:kappa}
is then obtained from the coefficient of the $q^2$ term in the Taylor
expansion of the inverse structure factor, which is related to the
free energy functional by \cite{binder83} $S^{-1}({\bf q}) = (k_BT)^{-1}
\delta^2\!F/\delta \phi^2$, where the r.h.s. is evaluated in q-space.

Because of the approximations made in the calculation of the square
gradient expression, the Flory-Huggins-de Gennes free energy
functional describes the physics of blends in the weak segregation
limit, and on length scales much larger than the average chain
dimension \cite{binder94,fredrickson92,tang92}. In
strongly-segregating blends ($\chi N \gg 1$) for which $\chi$ is small
but $N \to \infty$, Eq.~2.2 with a different prefactor in the
$\phi$-dependent part is typically used \cite{tang92,roe86,mcmullen92}.

The local part of the free energy (Eq.~\ref{eq:fh}) has the same
Ginzburg-Landau type of double-well structure as small molecules or
Ising-like systems \cite{goldenfeld92}.
Thus, the only difference between the free
energy functionals for the simplest small molecule and polymeric
systems arises from the chain connectivity, which is expressed in the
FHDG functional through the reduction of the entropic part of the
local term, and by the $\phi$-dependent part of the square-gradient
coefficient.  KM proposed that the entropic contribution to the
nonlocal part of the free energy, namely the concentration-dependent
square-gradient coefficient, provides the barrier to coarsening which,
when combined with the percolation-to-cluster transition, causes
pinning of the off-critical phase-separating blend.  In the next
section, we re-examine the numerical solution of the time evolution of
the CH-FHDG equation for both critical and off-critical blends.
Specifically, we show that for this model pinning is {\it not observed} in
the continuum limit, regardless of the blend composition,
although a dynamical exponent slightly smaller than $1/3$ is found.

The theoretical description of spinodal decomposition in binary blends
is based on the Cahn-Hilliard-Cook equation for the time evolution of
the concentration, originally derived for small
molecule systems \cite{cahn,cook}:
\begin{eqnarray}
{\partial \phi \over \partial t} = \nabla \cdot \left(M(\phi) \nabla
{\delta F\{\phi\} \over  \delta \phi}\right) + \eta({\bf r},t).
\label{eq:ch}
\end{eqnarray}
In this equation, $M(\phi)$ is the mobility, $F\{\phi\}$ is the
coarse-grained free energy functional, and $\eta$ is thermal noise.
For polymers, the free energy functional is typically taken to be of
the Flory-Huggins-De Gennes form in Eq.~\ref{eq:fhdg}, but more
general free energy functionals may be included.

In the following, we will always consider for simplicity a symmetric
blend, for which $N_A = N_B \equiv N$ and $\sigma_A = \sigma_B \equiv
\sigma$.  In this case, the mobility
\begin{eqnarray}
M(\phi) = N D \phi (1-\phi)
\label{eq:mob}
\end{eqnarray}
has been proposed \cite{degennes80}, where $D$ is the self-diffusion
coefficient. We will also take the effective interaction distance $\lambda$
equal to the Kuhn length $\sigma$.
Substituting in Eq.~\ref{eq:ch} the functional derivative of $F$ and
the expression for $M$, the time evolution of the concentration is
given by:
\begin{eqnarray}
{\partial \phi({\bf r}, t) \over \partial t} = N D
\nabla \left\{ \phi (1-\phi)
\nabla \left[ {1 \over N} \ln {\phi \over 1- \phi} + \chi (1- 2\phi) - \right.
\right. \cr \cr
\left(2 \chi \sigma^2 + {\sigma^2 \over 18 \phi (1-\phi)} \right) \nabla^2 \phi
\left. \left. + {(1-2 \phi) \sigma^2 \over 36 \phi^2 (1- \phi)^2} (\nabla
\phi)^2 \right]
\right\}.
\label{eq:chfhdg1}
\end{eqnarray}
Note that in Eq.~\ref{eq:chfhdg1} we have neglected the thermal noise
term. Since we are interested in the presence or absence of pinning
due to the FHDG free energy functional alone, and since it has been
shown that the presence of thermal noise in the analogue of this
equation for small molecule systems does not influence the scaling
function or the growth exponent during coarsening\cite{binder90,rogers88},
we will neglect noise in our simulations\cite{ctgm89,brown93}.

Eq.~\ref{eq:chfhdg1} can be rescaled so that it depends on dimensionless
space and time variables. The transformation, valid only in the
unstable region, is the following \cite{kotnis92}:
\begin{eqnarray}
{\bf x} = {(\chi - \chi_s)^{1/2} \over \sigma} {\bf r}
\hspace{1cm}
\tau = {D (\chi - \chi_s)^2 \over \sigma^2 \chi_s} t,
\end{eqnarray}
where $\chi_s = 1/(2N \phi_0 (1-\phi_0))$ gives the spinodal curve and
$\phi_0$ is the average value of concentration.
(This rescaling differs from the rescaling commonly used in experiments
by a simple numerical factor.)  After this transformation, Eq.~\ref{eq:chfhdg1}
becomes \cite{kotnis92}:
\begin{eqnarray}
{\partial \phi({\bf x}, \tau) \over \partial \tau} = {1 \over 2 \phi_0
(1-\phi_0)} \nabla \left\{ \phi (1-\phi)
\nabla \left[ {\chi_c \over 2(\chi - \chi_s)} \ln {\phi \over 1- \phi} - 2
{\chi
\over \chi - \chi_s} \phi - \right. \right. \cr \cr
\left. \left. \left(2 \chi \sigma^2 + {\sigma^2 \over 18 \phi (1-\phi)} \right)
\nabla^2 \phi
+ {(1-2 \phi) \sigma^2 \over 36 \phi^2 (1- \phi)^2} (\nabla \phi)^2 \right]
\right\}.
\label{eq:chfhdg2}
\end{eqnarray}

\section{Numerical Solution}
\label{sec:solution}

The solution of this continuum equation (~\ref{eq:chfhdg2}, which was first
studied by KM\cite{kotnis92}, describes the time evolution of the
concentration field after a quench to $\chi>\chi_s$ in the unstable
region \cite{ctgm89,brown93}.  The initial condition before the
quench, corresponding to high temperature, is given by a uniform field
with random fluctuations about its average value $\phi_0$.  At early
times following the quench, the uniform concentration is unstable with
respect to the long wavelength fluctuations arising from the initial
condition, and the two components begin to spatially separate.
Domains rich in one or the other component form, and then coarsen so
as to remove interfaces and minimize the free energy.  In
small-molecule mixtures described by, e.g., the Ginzburg-Landau free
energy functional, these domains coarsen until phase separation is
complete regardless of the relative composition and the presence of
thermal noise. Our goal in this paper is to determine if the same
is true for the FHDG free energy functional \cite{brown93}.

In the late stages of decomposition, the system can be characterized
by the evolution of the typical size of growing domains.  The scaling
hypothesis \cite{binder74} states that this length $L$ (as calculated
from, e.g., the inverse of the peak position of the structure factor,
the inverse of the first moment of the structure factor, the position
of the first zero in the pair correlation function, etc.) evolves in
time according to
\begin{eqnarray}
L \sim \tau^{\alpha}.
\end{eqnarray}
The choice of one particular definition of $L$ is dictated only by
convenience. Numerical simulations,
experiments, and analytical results strongly support the
validity of the scaling hypothesis in small molecule systems, giving
in the absence of hydrodynamic forces the value $\alpha = 1/3$
independent of quench depth, relative composition, and the presence of
noise\cite{binder90}. While polymers are believed to belong to the same static
universality class as small molecules, the situation is less clear
with respect to dynamics.

To study the kinetics of spinodal decomposition in polymer blends, we
numerically integrate Eq.~\ref{eq:chfhdg2} via a finite difference
scheme for both time and space variables.  The continuous space of
position vectors is replaced by $n^3$ sites on a simple cubic lattice
with mesh size (lattice spacing) $\Delta x$.  The temporal
discretization is achieved by replacing the continuous time variable
$\tau$ by a series of $m$ discrete time steps of duration $\Delta
\tau$.  The value of the concentration field at all sites at each time
step is then computed by a first-order Euler numerical integration
scheme \cite{numrecipes}, described in detail in the Appendix.

Although a large time step and mesh size would speed up the
computation, the mesh size must be chosen carefully so as to be
smaller than the smallest important length scale in the problem at all
times --- here the interfacial width, which decreases in time
until the latest stages of demixing. The size of the time step is in
turn limited by the mesh size.  A time step that is too large could
generate instabilities and spurious solutions \cite{rogers88,numrecipes}.
Thus, these discrete variables must be chosen carefully in concert.
A linear stability analysis can be helpful in suggesting reasonable
trial values.  If the algorithm is stable
with these values of $\Delta x$ and $\Delta \tau$, one can then vary
them to find optimum values and to ensure that the solution is
accurate and {\it independent} of the choice of these parameters.

We have studied the effect on the numerical solution of
Eq.~\ref{eq:chfhdg2} of changing $\Delta x$; $\Delta \tau$ is changed
suitably so as to maintain stability.  The boundary conditions are
periodic in all three directions.  Initial conditions are given by
random values of the concentration field, with average $\phi_0$ and a
flat distribution between $\phi_0 - \Delta
\phi_0$ and $\phi_0 + \Delta \phi_0$.  $\Delta \phi_0$ has a strong
influence on the behavior during the initial regime, but does not
affect the late stage behavior.  Therefore we fix $\Delta \phi_0 =
0.1$.  Several realizations of the initial conditions were averaged
together for every set of parameters.

The phase separation is monitored visually in real space, and
quantitatively by determining the time evolution of $k_1(\tau)$, the
first moment of the spherically-averaged structure factor
\cite{glotzer95}; this is the inverse length that is used to determine
the exponent $\alpha$.  For the purposes of comparison with previous
studies\cite{kotnis92,brown93}, we take $\chi$ to be related to $T$
(K) by \cite{wiltzius}:
\begin{eqnarray}
\chi = 0.326/T -2.3 \cdot 10^{-4},
\end{eqnarray}
and fix $T_c = 62\:^o C$. The system was quenched to temperatures
$T=54.5, 49$ and $25 \;^o C$ for critical composition ($\phi_o = 0.5$),
and to temperatures $T=35$ and $15 \; ^o C$ for off-critical composition
($\phi_o = 0.4$).

We first consider the solution of Eq.~\ref{eq:chfhdg2} obtained with a
mesh size $\Delta x = 1$, time step $\Delta \tau = 0.01$ and $n = 32$
(Fig.~\ref{n=32}).  This choice of mesh size gives, e.g. for $T=35 \; ^o
C$ and $\phi_0 = 0.4$, an equivalent dimensional mesh size of $\Delta
r \approx 6 R_g$, where $R_g$ is the average chain radius of gyration.
With these choices, we are exactly repeating the integration of
Eq.~\ref{eq:chfhdg2} previously performed by Kotnis and Muthukumar.
Our results reproduce their findings.  For critical quenches ($\phi_0=
0.5$), after an initial transient, the system enters the late stage
regime, where $k_1$ decays in time with a power law.  The exponent
$\alpha$ appears to be smaller than $1/3$, the value expected for
spinodal decomposition in small molecule systems and for polymer
blends in the intermediate stages of demixing (ie.~without
hydrodynamics).  For off-critical quenches ($\phi_0=0.4$), domain
growth stops before the phase separation is complete.

Results change drastically when the mesh size is reduced to $\Delta x
= 0.5$, for which the time step must be reduced to $\Delta \tau =
0.002$ to maintain numerical stability (Fig.~\ref{n=64}). (Note that
in this case we must take $n = 64$ to keep the system size the same as
before --- $n \cdot \Delta x = 32$).  This choice of mesh size gives,
e.g.  for $T=35 \;^o C$ and $\phi_0 = 0.4$, an equivalent dimensional
mesh size of $\Delta r \approx 3 R_g$.  In this case, the late stage
behavior is the same for both critical and off-critical quenches, even
after the blend undergoes the percolation-to-cluster transition.
After an initial transient, the late stage scaling regime is reached:
$k_1$ decays as a power law and no pinning is observed. The values of
$\alpha$ can be computed from the slopes of the curves in
Fig.~\ref{n=64} and are reported in Table~\ref{tab1}.  In all cases
$\alpha$ is found to be greater than $1/4$ and smaller than $1/3$.  We
believe that the latter is the true asymptotic value for this free
energy functional; the small systematic error may be attributed to the
crossover from the preasymptotic regime and, possibly, to the
numerical slowing down given by a still oversized mesh size (see
Sec.~\ref{sec:discussion}).  Increased accuracy might be obtained by
using an even smaller value for $\Delta x$, but this implies a
suitable reduction in the value of $\Delta \tau$, and as a result, a
prohibitively large computation time.

The critical relevance of the mesh size to the late stage behavior of
the system is also evident from Fig.~\ref{comparison}, where a
``pinned'' state obtained with $\Delta x = 1$ is ``depinned'' by
reducing the mesh size to $0.5$.

\section{Discussion}
\label{sec:discussion}

The absence of pinning for small mesh sizes in our simulations even
for off-critical quenches clearly shows that what is observed for
$\Delta x = 1$ is not a physical effect but only an artifact of the
discretization.  By integrating Eq.~\ref{eq:chfhdg2} via a
discretization scheme, we are actually changing the model under
consideration: the solution of the discrete model exactly reproduces
that of the continuum equation only in the limit where $\Delta t$ and
$\Delta x$ approach zero.  For this reason one must always confirm
that the numerical results are independent of the values of the
discretization variables.  In particular, it was recently shown
that an oversized mesh size can cause a
non-physical freezing of interfacial motion for systems with
non-conserved order parameter\cite{osher94}.  ``Spurious pinning'' was
already noted by Rogers et al. \cite{rogers88}
in a conserved order parameter system, who showed that
for the Cahn-Hilliard equation with the Ginzburg-Landau free energy
functional, a mesh size $\Delta x >1.7$ causes an unphysical decrease
in the effective growth exponent.

In general, the solution of the discrete model should reproduce the
behavior of the continuum equation if $\Delta x$ is much less than the
smallest physical length modeled in the system.  In the case under
investigation, we are studying a polymer blend in the weak segregation
limit ($\chi \ge \chi_c, \chi \ll 1$).  The smallest physical length
that must be resolved is the interfacial width, which during the late
stages of phase separation is of the order of the correlation length $\xi$
\cite{fredrickson92}:
\begin{eqnarray}
\xi \sim {R_g \over (\chi - \chi_c)^{1/2} \cdot N^{1/2}} \sim {\sigma \over
(\chi - \chi_c)^{1/2}},
\end{eqnarray}
which is close to unity in rescaled units. Choosing $\Delta x = 1$
implies that we are resolving the system at a length {\it equal} to
the correlation length. The spatial derivatives at the interfaces are
consequently computed inaccurately with this mesh size, thereby
producing unphysical results.  This picture is confirmed by
Fig.~\ref{profile} showing that when $\Delta x = 1$ the interface
appears only one mesh size wide.  When $\Delta x = 0.5$ the interface
is smoother, and larger than the mesh size; hence no pinning occurs.  This
effect also explains the low estimates of $\alpha$ for critical quenches
when $\Delta x =1$ ---
the sharpness of the interface unphysically slows down the evolution
of the solution, even if it is not sufficient to pin it.  Possibly
also the results for $\Delta x = 0.5$ are slightly biased by this
effect.

It is possible to understand the origin of this problem in another way
by looking directly at the equation of motion.  Consider for
simplicity a one-dimensional small molecule system (i.e. the
square-gradient coefficient $\kappa$ and the mobility $M$ are constant),
where the concentration profile goes from one bulk value ($\phi_1$) at
site $x_i-\Delta x$ to the other ($\phi_2$) at site $x_i+\Delta x$, through an
interface.
For the domain size to grow, the interface must move, and thus $\phi(x_i)$ must
change from $\phi_1$ to $\phi_2$.  The driving force for this change
is the square gradient term in the free energy, which yields a
Laplacian in the functional derivative of $F$.  This force must
overcome the double well potential given by the local term in the free
energy expression, as stated by the Cahn-Hilliard equation:
\begin{eqnarray}
{\partial \phi \over \partial t} = M \nabla^2 \left({\partial f \over \partial
\phi} - \kappa \nabla^2 \phi \right).
\end{eqnarray}
In the discrete version of this equation, the local part does not
depend on $\Delta x$, while the Laplacian is given by:
\begin{eqnarray}
\nabla^2 \phi = {1 \over (\Delta x) ^2} \left[ \phi(x_i+\Delta x)+
\phi(x_i-\Delta x) - 2  \phi(x_i) \right].
\end{eqnarray}
When we increase $\Delta x$ the denominator grows indefinitely, while
the numerator is bounded above by $\phi_2+\phi_1- 2 \phi(x_i)=
\mbox{const}$.  Then, increasing $\Delta x$
{\it decreases} the value of the Laplacian term, while the local term
is unchanged.  For $\Delta x$ large enough
the Laplacian cannot overcome the local term, the solution stops
evolving, and the system artificially ``pins''.

\section{Conclusion}
\label{sec:conclusion}

In summary, we have shown that both critical and off-critical polymer
blends described by the Flory-Huggins-De Gennes free energy functional
undergo spinodal decomposition via the Cahn-Hilliard equation without
pinning of the domain growth as observed in some experiments. Even in
the absence of thermal noise\cite{brown93}, the solution of the
discretized equation of motion shows coarsening without evidence of
pinning, regardless of the relative concentration of the blend
components. We have also shown that previous solutions of the CH-FHDG
equation that exhibited pinning were artifacts of an oversized mesh
size used in the discretization and numerical integration of the
equation of motion, and {\it not} a result of the
concentration-dependence of the square gradient coefficient as
previously suggested \cite{kotnis92}.  This suggests the FHDG free
energy functional alone, as written in Eq.~\ref{eq:fhdg} is not sufficient
to describe the physics responsible for pinning in real blends.

Evidently, a model able to describe the arrested growth observed in
experiments must include additional physical ingredients.
One should consider that experimental blends which exhibit pinning often
contain components which are not simple homopolymers; such is the case with
Hashimoto's random copolymer/homopolymer blends, as well as with
Hasegawa's liquid-crystalline polymer/homopolymer blend, in which the
liquid-crystalline component is anistropic at the quench temperatures
at which pinning was observed. These blends may not be describable by
the simple Flory-Huggins-De Gennes expression, and consequently we
should not expect the CH-FHDG equation to mimic their behavior during
spinodal decomposition. It is
also important to note that a significant fraction of homopolymer
blends never exhibit pinning, regardless of the relative composition.
However, for those that do, it is possible that either the mobility,
or free energy functional, or both, must be appropriately modified.
The simplified expression for mobility (Eq.~\ref{eq:mob}) used in the
CH-FHDG equation, which was originally derived only for the special
case of a perfectly symmetric blend \cite{degennes80}, may neglect
important contributions to mobility arising from the connectivity
within the polymer chains.

With respect to possible free energy
modifications, it is well known that interfacial growth can be slowed
or stopped by decreasing the interfacial tension. This can be achieved by,
e.g. using surfactants in the case of small molecules\cite{surfactants},
diblock copolymers in the case of polymers\cite{copolymers}, or impurities
in the case of alloys\cite{impurities}.
If the experimental blends contain even a small
number of impurities, or specific interaction regions that act as
impurities \cite{ggscs94}, these impurities could arrest the demixing
process and cause pinning.  Finally, the FHDG free energy functional
in Eq.~\ref{eq:fhdg} describes incompressible blends; real blends are
in fact compressible. A coupling of concentration fluctuations and
density fluctuations may be responsible for pinning in blends
\cite{douglas}, in which case a reformulation of the free energy
functional as well as the addition of a second order
parameter field is necessary.

We are extremely grateful to J. Douglas, B. Hammouda, P. Gallagher, C.
Han, E. DiMarzio, G. McFadden, A. Coniglio, F. Corberi and J. Warren,
and especially to M. Muthukumar, A. Chakrabarti and G. Brown, for
useful discussions.
We thank the Center for Computational Science at Boston University and
the University of Maryland for generous use of their CM-5.
CC would like to thank the Structure and Mechanics
Group in the Polymers Division at NIST, and the NIST Center for
Theoretical and Computational Materials Science, for their
hospitality.

\section{Appendix}

The numerical solution of Eq.~\ref{eq:chfhdg2} is performed via iteration
of the following map:
\begin{eqnarray}
\phi^{m+1}_{i,j,k} = \phi^m_{i,j,k} +
\Delta \tau {\partial \phi^m_{i,j,k} \over \partial \tau}.
\end{eqnarray}
This map, given the value of the concentration field $\phi^m_{i,j,k}$ at
time $m \Delta \tau$ at each of the $n^3$ sites of a simple cubic
lattice with mesh size $\Delta x$, yields the value of $\phi^{m+1}_{i,j,k}$ at
each site at time $(m+1) \Delta \tau$. Note that the mesh size is
taken to be the same in all directions.

The time derivative $\partial \phi^m_{i,j,k} / \partial \tau$ is given
by the discretization of the left hand side of Eq.~\ref{eq:chfhdg2},
with spatial derivatives centrally discretized.  This means the
chemical potential $\mu^m_{i,j,k}$ that appears in square brackets in
Eq.~\ref{eq:chfhdg2} is computed using:
\begin{eqnarray}
\left[\nabla \phi({\bf x},\tau)\right]^2 \to
\left( {\phi^m_{i+1,j,k} - \phi^m_{i-1,j,k} \over 2\Delta x}\right)^2 +
\left( {\phi^m_{i,j+1,k} - \phi^m_{i,j-1,k} \over 2\Delta x}\right)^2 +
\left( {\phi^m_{i,j,k+1} - \phi^m_{i,j,k-1} \over 2\Delta x}\right)^2,
\end{eqnarray}
and
\begin{eqnarray}
\nabla^2 \phi({\bf x},\tau) \to
{\phi^m_{i+1,j,k} - 2\phi^m_{i,j,k} + \phi^m_{i-1,j,k} \over (\Delta x)^2} +
{\phi^m_{i,j+1,k} - 2\phi^m_{i,j,k} + \phi^m_{i,j-1,k} \over (\Delta x)^2} +
\cr \cr
{\phi^m_{i,j,k+1} - 2\phi^m_{i,j,k} + \phi^m_{i,j,k-1} \over (\Delta x)^2}.
\end{eqnarray}

The divergence of the product $\phi (1-\phi) \nabla \mu$ is evaluated
as follows:
\begin{eqnarray}
\nabla \cdot \{ \phi (1-\phi) \nabla \mu \} \to
\left( {X^m_{i+1,j,k} - X^m_{i-1,j,k} \over 2\Delta x} \right) +
\left( {Y^m_{i,j+1,k} - Y^m_{i,j-1,k} \over 2\Delta x} \right) + \cr \cr
\left( {Z^m_{i,j,k+1} - Z^m_{i,j,k-1} \over 2\Delta x} \right),
\end{eqnarray}
where
\begin{eqnarray}
X^m_{i,j,k} =
\left( {\mu^m_{i+1,j,k} - \mu^m_{i-1,j,k} \over 2\Delta x} \right)
\left( \phi^m_{i,j,k} (1 - \phi^m_{i,j,k}) \right),
\end{eqnarray}
and $Y^m_{i,j,k}$ and $Z^m_{i,j,k}$ are defined accordingly.

\begin{figure}
\caption{Plot of $\log(k_1)$ vs $\log(\tau)$ for a system with $n=32$ and
$\Delta x = 1$. Each curve is the average over 5 realizations with different
initial conditions.}
\label{n=32}
\end{figure}

\begin{figure}
\caption{Plot of $\log(k_1)$ vs $log(\tau)$ for a system with $n=64$ and
$\Delta x = 0.5$. Each curve for critical quenches is the average over 5
realizations with different initial conditions. Each curve for off-critical
quenches is averaged over 10 realizations.}
\label{n=64}
\end{figure}

\begin{figure}
\caption{(a) Snapshot of a 2D slice of a system of size $n=32$ and
$\Delta x = 1$ at $\tau = 100$. (b) The same system for $\tau = 200$.
(c) The same system in (a) at $\tau=200$ when $\Delta x$ is switched to $0.5$
at $\tau = 100$. The decrease of mesh size allows the system to phase-separate
without pinning.}
\label{comparison}
\end{figure}

\begin{figure}
\caption{Interface profile for (a) $\Delta x = 1$ and (b) $\Delta x = 0.5$,
for $n = 32$, $\phi_0 = 0.4$, $T = 15$ and $\tau = 100$.}
\label{profile}
\end{figure}

\begin{table}
\centering
\begin{tabular}{lccc}
\multicolumn{4} {c}{Critical quenches ($\Delta \phi_o = 0.5$)}  \\ \smallskip
T        & 25               & 49                & 54.5              \\
\smallskip
$\alpha$ &$0.263 \pm 0.001$ &$0.286 \pm 0.002$&$0.270 \pm 0.001$    \\
\end{tabular}
\medskip
\begin{tabular}{lcc}
\multicolumn{3} {c}{Off-critical quenches ($\Delta \phi_o = 0.4$)}\\ \smallskip
T        & 15               & 35                    \\ \smallskip
$\alpha$ &$0.290 \pm 0.001$ &$0.306 \pm 0.001$      \\
\end {tabular}

\caption{Values of the dynamic exponent $\alpha$ for $n = 64$ and
$\Delta x = 0.5$, computed for $\tau > 10$. The exponents and errors
were computed using a linear regression fit of the average values
plotted in Fig. 2.}
\label{tab1}
\end{table}

%\end{narrowtext}


\begin{thebibliography}{99}

\bibitem{hasegawa88} H. Hasegawa, T. Shiwaku, A. Nakai and T. Hashimoto, in
{\it Dynamics of Ordering Processes in Condensed Matter}, edited by S. Komura
and H. Furukawa (Plenum, New York,
1988).

\bibitem{hashimoto88} T. Hashimoto, in
{\it Dynamics of Ordering Processes in Condensed Matter}, edited by S. Komura
and H. Furukawa (Plenum, New York,
1988).

\bibitem{wong} A. Wong and P. Wiltzius (unpublished).

\bibitem{hashimoto92} T. Hashimoto, M. Takenaka, and T. Izumitani,
J. Chem. Phys. {\bf 97}, 679 (1992); M. Takenaka, T. Izumitani and
T. Hashimoto, J. Chem. Phys. {\bf 98}, 3528 (1993).

\bibitem{hashimoto94} T. Hashimoto in
{\it Material Science and Technology: Phase Transformations in
Materials \/} {\bf 12}, edited by P. Haasen (Weinham: VCH, 1993), 251.

\bibitem{lauger94} J. L\"auger, R. Lay,
and W. Gronski, J. Chem. Phys.  {\bf 101}, 7181 (1994).

\bibitem{cahn} J.W. Cahn and J.E. Hilliard, J. Chem. Phys. {\bf 28},
258 (1958); J.W. Cahn, Acta. Metall. {\bf 9}, 795 (1961); Acta
Metall. {\bf 10}, 179 (1962); J. Applied Phys. {\bf 34}, 3581 (1963);
General Electric Research Lab. Report, RL3561M (1964); Acta
Metall. {\bf 14}, 1685 (1966); Trans. Metall. Soc. AIME {\bf 242}, 166
(1968).

\bibitem{gunton83} J.D.~Gunton, M. San Miguel and P.S.~Sahni, in {\it Phase
Transitions and Critical Phenomena\/} {\bf 8}, edited by C. Domb and
J.  L.~Lebowitz (Academic Press, London, 1983).

\bibitem{hashimotopt} T. Hashimoto, Phase Transitions {\bf 12}, 47 (1988).

\bibitem{binder90} K. Binder in
{\it Material Science and Technology: Phase Transformations in
Materials \/} {\bf 5}, edited by P. Haasen (Weinham: VCH, 1990),
pp.~405-471.

\bibitem{han92} C.C. Han and A.Z. Akcasu, Ann. Rev. Phys. Chem. {\bf 43}, 61
(1992).

\bibitem{binder94} K. Binder, Adv. in Poly. Sci. {\bf 112}, 181 (1994).

\bibitem{glotzer95} S.C. Glotzer in {\it Annual Reviews of
 Computational Physics}, {\bf II}, 1, D. Stauffer, ed. (World
 Scientific, Singapore, 1995).

\bibitem{hayward87} S. Hayward, D.W. Heermann, and K. Binder,
J. Stat. Phys. {\bf 49},  1053 (1987).

\bibitem{kotnis92} M. Kotnis and M. Muthukumar, Macromolecules {\bf 25}, 1716
(1992).

\bibitem{helfand} E. Helfand and Y. Tagami, J. Chem. Phys. {\bf 56}, 3592
(1972);
E. Helfand and A.M. Sapse, J. Chem. Phys. {\bf 62}, 1327 (1975);
E. Helfand, Macromolecules {\bf 9}, 307 (1976).

\bibitem{lifshitz61} I.M.~Lifshitz and V.V.~Slyozov, J.~Phys.~Chem., Solids
{\bf 19}, 35 (1961).

\bibitem{huse86} D.A. Huse, Phys. Rev. B {\bf 34}, 7845 (1986).
\bibitem{degennes80}
P.G. De Gennes, J. Phys.(Paris) Lett. {\bf 38L}, 441 (1977).
P.G. De Gennes, J. Chem. Phys. {\bf 72}, 4756 (1980).

\bibitem{pincus81} P. Pincus, J. Chem. Phys. {\bf 75}, 1996 (1981).

\bibitem{binder83} K. Binder, J. Chem. Phys. {\bf 79}, 6387 (1983).

\bibitem{flory} P.J. Flory, J. Chem. Phys. {\bf 9}, 660 (1941);
{\it Principles of Polymer Chemistry}, Cornell University Press, Ithaca, 1953.

\bibitem{huggins} M.L. Huggins, J. Chem. Phys. {\bf 9}, 440 (1941).

\bibitem{fredrickson92} For a nice discussion, see
G. Fredrickson in {\it Physics of Polymer Surfaces
and Interfaces}, edited by I.C.~Sanchez, (Butterworth-Heinemann,
Boston, 1992).

\bibitem{edwards66} S.F. Edwards, Proc. Phys. Soc. {\bf 88}, 265 (1966).

\bibitem{akcasu}
A.Z. Akcasu and M. Tombakoglu, Macromolecules {\bf 23}, 607 (1989);
A.Z. Akcasu, R. Klein, and B. Hammouda, Macromolecules {\bf 26}, 4136 (1993).

\bibitem{tang92} H. Tang and K.F. Freed, J. Chem. Phys. {\bf 94}, 1572 (1991);
H. Tang and K.F. Freed, J. Chem. Phys. {\bf 94}, 6307 (1991);
X.C. Zeng, D.W. Oxtoby, H. Tang, and K.F. Freed, J. Chem. Phys. {\bf 96}, 4816
(1992).

\bibitem{roe86} R.J. Roe, Macromolecules {\bf 19}, 728 (1986).

\bibitem{mcmullen92} W. McMullen, in {\it Physics of Polymer Surfaces
and Interfaces}, edited by I.C.~Sanchez, (Butterworth-Heinemann,
Boston, 1992).

\bibitem{goldenfeld92} N. Goldenfeld, {\it Lectures on Phase Transitions and
the Renormalization Group}, Addison-Wesley Publication Co., MA, 1992.

\bibitem{cook} H.E. Cook, Acta Metall. {\bf 18}, 297 (1970).

\bibitem{rogers88} T.M. Rogers, K.R. Elder and R.C. Desai, Phys.~Rev.~B {\bf
37}, 9638 (1988), and references therein.

\bibitem{ctgm89} Eq.~\ref{eq:chfhdg1}
with concentration-independent mobility and thermal noise was first
studied for critical quenches in A. Chakrabarti, R. Toral, J.D. Gunton
and M.  Muthukumar, Phys. Rev. Lett. {\bf 63}, 2072 (1989); A.
Chakrabarti, R.  Toral, J.D. Gunton and M. Muthukumar, J. Chem. Phys.
{\bf 92}, 6899 (1990); J.D.Gunton, R. Toral and A. Chakrabarti,
Physica Scripta {\bf T 33}, 12 (1990). Solutions of that equation and
the CH-GL equation were compared for critical quenches in A.
Chakrabarti and G. Brown, Phys. Rev.  A {\bf 46}, 981 (1992).

\bibitem{brown93} Eq.~\ref{eq:chfhdg1} with concentration-independent mobility
and thermal noise was studied for off-critical quenches in G. Brown and
A. Chakrabarti, J. Chem. Phys. {\bf 98}, 2451 (1993). No pinning was observed
in their simulations.

\bibitem{binder74}
K. Binder and D. Stauffer, Phys. Rev. Lett. {\bf 33}, 1006 (1974).

\bibitem{numrecipes} W.H. Press, B.P. Flannery, S.A. Teukolsky and
W.T. Vetterling, {\it Numerical Recipes}, Cambridge University Press,
Cambridge, 1990.

\bibitem{wiltzius} P. Wiltzius, F.S. Bates and W. Heffner, Phys. Rev. Lett.
{\bf 60}, 1538 (1988); F.S. Bates and P. Wiltzius, J. Chem. Phys. {\bf 91},
3258 (1989).

\bibitem{osher94} B. Merriman, J.K. Bence and S.J. Osher, J. Comp. Phys.
{\bf 112}, 334 (1994).

\bibitem{surfactants} Gompper and Schick, in {\it Phase Transitions and
Critical
Phenomena} {\bf 16}, Domb and Lebowitz, eds. 1994;
M. Laradji, H. Guo, M. Grant, and M.J. Zuckermann, J. Phys. A {\bf 24}, L629
(1991);
J. Phys. Condens. Matter {\bf 4}, 6715 (1992).

\bibitem{copolymers}
T. Hashimoto and T. Izumitani, Macromolecules {\bf 26}, 3631 (1992).

\bibitem{impurities}
D.A. Huse and C.H. Henley, Phys. Rev. Lett. {\bf 54}, 2708 (1985);
G. Grest and D.J. Srolovitz, Phys. Rev. B {\bf 32}, 3014 (1985);
D.J. Srolovitz and G. Grest, Phys. Rev. B {\bf 32}, 3021 (1985);
S. Puri, D. Chowdhury and N. Parekh, J. Phys. A {\bf 24}, L1087 (1991);
S. Puri and N. Parekh, J. Phys. A {\bf 25}, 4127 (1992).

\bibitem{ggscs94} S.C. Glotzer, M.F. Gyure, F. Sciortino, A. Coniglio, and
H.E. Stanley, Phys. Rev. Lett. {\bf 70}, 3275 (1993); Phys. Rev. E
{\bf 49}, 247 (1994); F. Sciortino, P. Alstrom, R. Bansil, and H.E.
Stanley, Phys. Rev. E {\bf 47}, 4615 (1993).

\bibitem{douglas} J.F. Douglas, private communication.

\end{thebibliography}
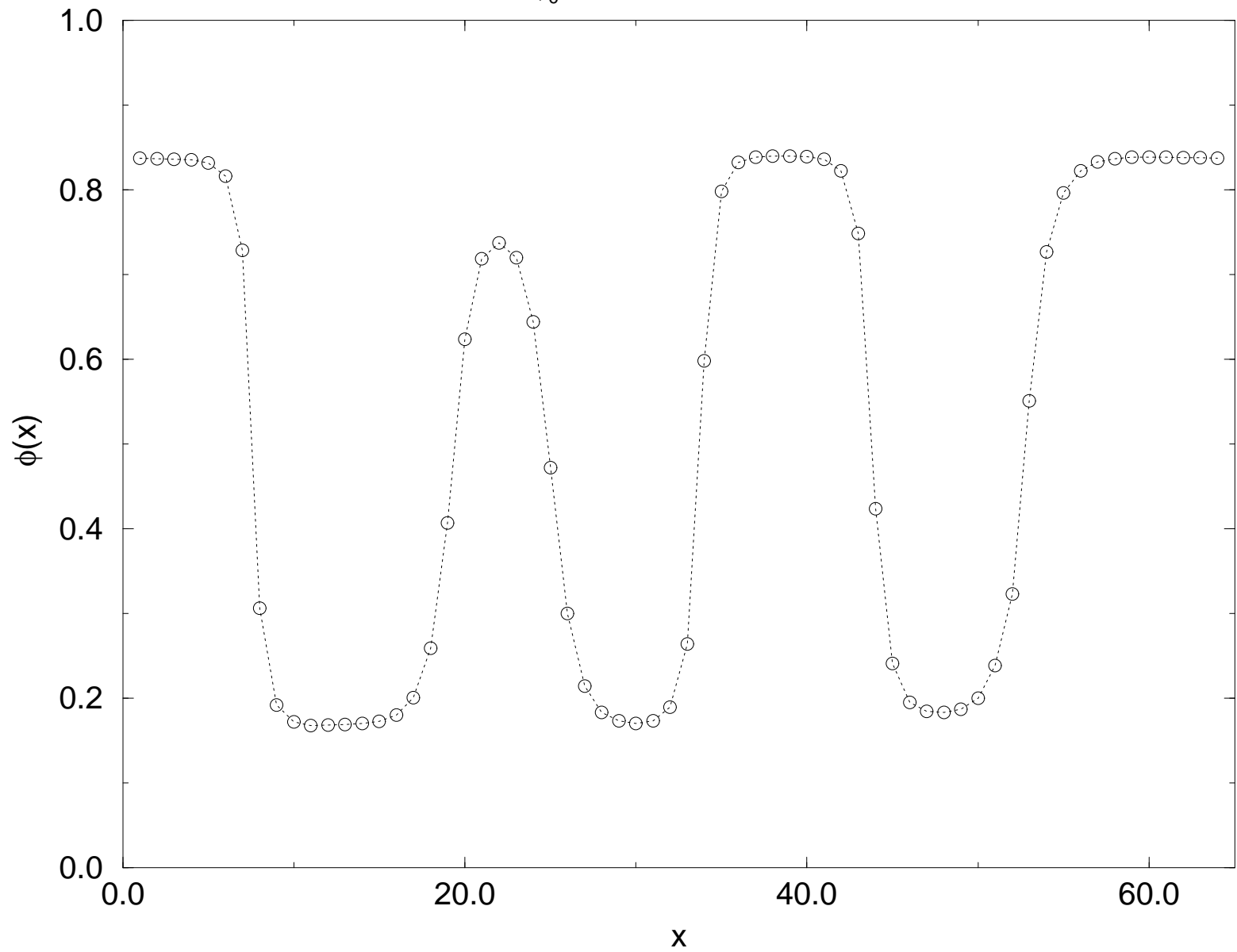
\end{document}